\documentclass[onecolumn,prc,superscriptaddress,unsortedaddress,preprintnumbers,amsmath,amssymb]{revtex4}

\usepackage{graphicx}
\usepackage{dcolumn}
\usepackage{bm}
\usepackage{txfonts}
\usepackage{bm}

\def\beq{\begin{equation}}
\def\eeq{\end{equation}}
\def\bea{\begin{eqnarray}}
\def\eea{\end{eqnarray}}

\def\fun#1#2{\lower3.6pt\vbox{\baselineskip0pt\lineskip.9pt
  \ialign{$\mathsurround=0pt#1\hfil##\hfil$\crcr#2\crcr\sim\crcr}}}
  
\begin{document}
\preprint{}

\title{Role of nucleonic Fermi surface depletion in neutron star cooling}

\author{J. M. Dong}\affiliation{Institute of Modern Physics, Chinese
Academy of Sciences, Lanzhou 730000, China}
\author{U. Lombardo}\affiliation{Universita di Catania and Laboratori Nazionali del Sud (INFN), Catania 95123, Italy}
\author{H. F. Zhang}\affiliation{School of Nuclear Science and Technology, Lanzhou University, Lanzhou 730000, China}
\author{W. Zuo}
\affiliation{Institute of Modern Physics, Chinese Academy of
Sciences, Lanzhou 730000, China}

\date{\today}

\begin{abstract}
The Fermi surface depletion of beta-stable nuclear matter is calculated
to study its effects on several physical properties which determine the
neutron star thermal evolution. The neutron and proton $Z$ factors measuring
the corresponding Fermi surface depletions, are calculated within the Brueckner-Hartree-Fock
approach employing the AV18 two-body force supplemented by a microscopic three body force.
Neutrino emissivity, heat capacity and, in particular, neutron $^3PF_2$ superfluidity turn out
to be reduced, especially at high baryonic density, to such an extent that the cooling rates of young neutron stars are significantly slowed.

\end{abstract}

\maketitle
\noindent{} Online-only material: color figures
\maketitle

\section{Introduction}\label{intro}\noindent

Neutron stars (NSs), with typical masses $M\sim 1.4M_{\odot }$ and
radii $R\sim 10$km, are natural laboratories for investigating
exotic phenomena that lie outside the realm of terrestrial
laboratories. They have been arousing tremendous interest since they are
related to many branches of contemporary physics as well as
astronomy (Haensel et al. 2006). In particular, NSs play the role
of connecting  nuclear physics with astrophysics,
since they realize one of the densest forms of nuclear matter
in the observable universe. Concerning NSs, important astrophysical quantities
can be measured with increasing accuracy, such as mass, radius, surface
temperature, spin period and spin-down, which
provide valuable information and knowledge about these objects.
On the theoretical side, much attention has been paid to the
equation of state (EOS) of dense matter, including superfluid states,
and cooling mechanisms via neutrino emission to understand the NS
thermal evolution (Haensel et al. 2006).

Nowadays, the measurements on the NS surface
temperature such as Cas A (Heinke \& Ho 2010, Posselt et al. 2013), allow us to investigate the
thermal evolution of NSs more deeply (Page et al. 2011; Shternin et al. 2011; Blaschke et al. 2012, 2013;
Sedrakian 2013; Newton et al. 2013; Bonanno et al. 2014). Since the
thermal evolution is quite sensitive to the equation of state (EoS) of NS matter,
in particular to  its composition and superfluidy states, one may grasp the crucial
information and knowledge on the NS interior from this study. Thus, the
exploration of the NS cooling might solve some difficult
issues in nuclear physics, for instance, the density dependence of
symmetry energy at supranuclear densities. The NS cools down
via neutrino emission from the stellar interior in the first $10^5$
years (Yakovlev \& Pethick 2004), and several types of neutrino sources
in the NS cores have been proposed as cooling mechanisms, such as the direct Urca (DU),
modified Urca (MU) processes and nucleon-nucleon bremsstrahlung (see the review
articles, Yakovlev et al. 1999, 2001; Page et al. 2004, 2006).

It has been stressed that the NS cooling
relies on many factors (Yakovlev et al. 1999), including the neutrino emission mechanisms, heat capacity,
thermal conductivity and reheating mechanisms in dense and superfluid states of matter.
The latter has to be described as a quasi-degenerate Fermi system characterized
by a large depletion of the Fermi surface due to
the strong short-range correlations of nucleon-nucleon ($NN$) interaction (Migdal 1967). The $Z$-factor, that measures
the deviation from a perfect Fermi gas described by the Fermi-Dirac distribution ($Z=1$) can take values much less than one.
This fact influences the level density of nucleons around the Fermi surface that controls many properties
of fermion systems related to particle-hole excitations around the Fermi energy.
In our previous work, the $Z$ factor effect on the $^3PF_2$
superfluidity of pure neutron matter was studied (Dong et al. 2013) and found that
the gap was dramatically reduced. In this work, we generalize the previous approach to investigate the neutron
$^3PF_2$ superfluidity of asymmetric nuclear matter and $\beta$-stable matter.
The NS interior is assumed to be made of $npe\mu$ matter without exotic degrees of freedom.
In such a context the $Z$ factor effects on various neutrino processes and heat capacity,
and on NS cooling are calculated. The material is organized as follows. In Sec. II, superfluidity, various
neutrino processes and heat capacity affected by the Fermi surface depletion are
respectively calculated and analyzed. Based on those results as inputs, the
NS cooling is discussed in Sec. III. Finally a summary is given in Sec. IV.

\section{Quasi-degenerate nuclear matter in beta-stable state: superfluidity, neutrino emissivity and heat capacity}\label{intro}\noindent

\begin{figure}[htbp]
\begin{center}
\includegraphics[width=0.4\textwidth]{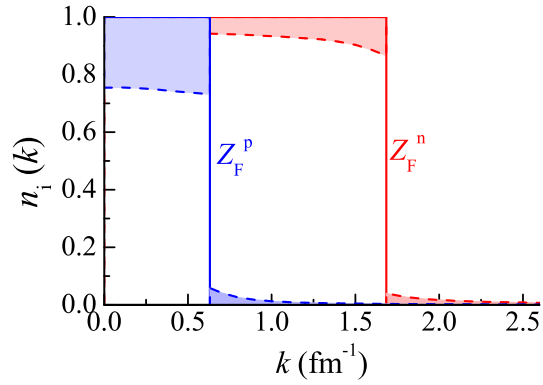}
\caption{Occupation probability of nucleons vs. momentum for
asymmetric matter with isospin asymmetry $\beta=0.9$ and density
$\rho=0.17$ fm$^{-3}$. (A color version of this figure is available
in the online journal.)}\label{fig:occupation}
\end{center}
\end{figure}

\begin{figure}[htbp]
\begin{center}
\includegraphics[width=0.45\textwidth]{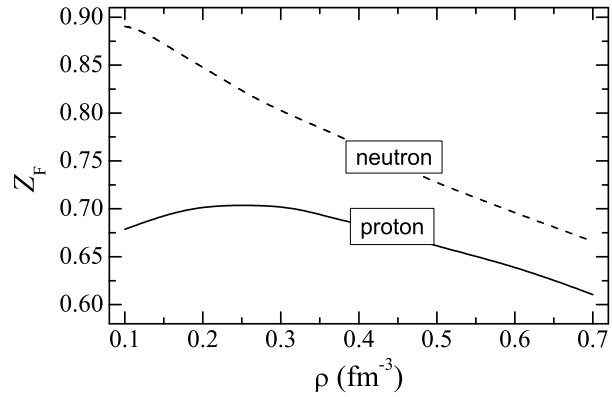}
\caption{$Z$ factors vs density in $\beta$-stable matter. The
fraction of each component is determined by symmetry energy from the
BHF approximation.}\label{fig:Z}
\end{center}
\end{figure}

The deviation of a correlated Fermi system from the ideal
degenerate Fermi gas is measured by the quasiparticle strength (Migdal 1967)
\begin{equation}
Z(k) = \left[ 1-\frac{\partial\Sigma(k,\omega)}{\partial\omega}\right]_{\omega=\epsilon(k)}^{-1},
\end{equation}
where $\Sigma(k,\omega)$ is the
self-energy as a function of momentum $k$ and energy $\omega$.
According to the Migdal-Luttinger theorem (Migdal 1960), the $Z$
factor at the Fermi surface equals the discontinuity of the
occupation number at the Fermi surface. We calculated the $Z$ factors
in the framework of the Brueckner-Hartree-Fock (BHF) approach by employing
the AV18 two-body force with a microscopic three body
force (Li et al. 2008). The self-energy is truncated to the fourth order of the expansion in powers of  G-matrix,
namely, $\Sigma =\Sigma _{1}+\Sigma _{2}+\Sigma _{3}+\Sigma _{4}$ (Jeukenne et al. 1976).
Figure~\ref{fig:occupation} displays, for the sake of illustration, the occupation probability
of neutrons and protons in strongly asymmetric nuclear matter. The momentum distribution around the
Fermi surface significantly departs from the typical profile of a
degenerate Fermi system especially at high densities, which is
attributed to the strong short-range correlations. In Figure~\ref{fig:Z}, we show the  $Z$ factors at the
Fermi surface $Z_F$ calculated for nuclear matter in a $\beta$-stable state, suitable for application
to NSs. The fraction of each component is determined by the EOS of nuclear matter from a BHF calculation.

In $\beta$-stable nuclear matter, the effect of the neutron-proton $^3$SD$_1$ coupling tensor channel
(namely, $I=0$ SD channel where $I$ is the total isospin of two nucleons) of the nuclear interaction
drives the deviation of neutron and proton Z-factors vs. total density. At very low density the system is mainly
composed of neutrons and the neutron $Z_F$  is that of a degenerate Fermi gas weakly interacting with few protons.
On the contrary, the very diluted proton fraction is strongly interacting, via the $I=0$ force, with large neutron excess,
resulting into a strong depletion of the proton momentum distribution at low density. As the total density increases, also
the proton fraction density increases for the competition with $I=1$ force inducing an increasing $Z_F$ value. Therefore, from the
interplay between the two mechanisms the  proton $Z_F$  first increases and then decreases as shown in the figure. The
calculated $Z$ factor will be used in the following calculations.


\subsection{Neutron $^3PF_2$ superfluidity}\label{intro}\noindent

The superfluidity gap at the Fermi surface quenches all processes that involve
elementary excitations around the Fermi surface, which could lead to an remarkable effect on
the NS cooling. The $^3PF_2$ superfluidity in pure neutron matter has been investigated in
a previous work with the inclusion of the $Z$ factor (Dong et al. 2013). Here we generalized
the investigation to isospin-asymmetric nuclear matter in $\beta$-stable condition. The $^3PF_2$
pairing gaps are determined by the following coupled equations
\begin{eqnarray}
\left( \begin{array}{l} \Delta_L ( k)\\ \Delta_{L+2}( k) \end{array} \right)
 = -\frac{1}{\pi}\int_0^\infty \!  k'^2{\rm d} k'  \frac{Z(k) Z(k')}{E_{k'}}\hspace*{1.5cm}\nonumber\\
\times \left( \begin{array}{ll} V_{L,L}(k,k') & V_{L,L+2}(k,k') \\ V_{L+2,L} (k,k') &
  V_{L+2,L+2}(k,k')
 \end{array} \right)
 \left(\begin{array}{l} \Delta_L  (k') \\ \Delta_{L+2} (k')\end{array} \right),\ \
\end{eqnarray}
with $E_{k}=\sqrt{[\epsilon (k)-\mu ]^{2}+\Delta _{k}^{2}}$ and
$\Delta =\sqrt{\Delta _{L}^{2}+\Delta _{L+2}^{2}}$.  $V_{L,L'}(k,k')$ is the
matrix element of the realistic $NN$ interaction including three-body forces. Here the angle-average approximation is adopted, and the relation
between the gap at $T=0$ and the critical temperature $T_c$ is given by $k_B T_c=0.57\Delta (T=0)$ (Baldo et al. 1992).

\begin{figure}[htbp]
\begin{center}
\includegraphics[width=0.45\textwidth]{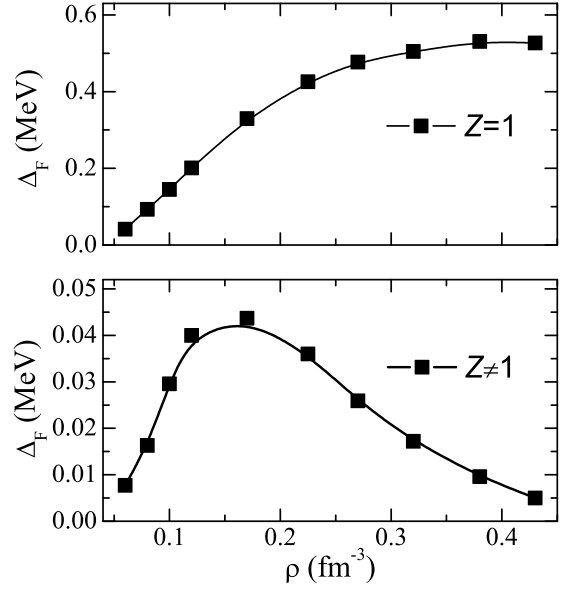}
\caption{Neutron $^3PF_2$ gap vs. total baryonic density in $\beta$-stable nuclear matter. The calculations with $Z=1$ and  $Z \neq 1$
are shown for comparison.}\label{fig:n3PF2-beta}
\end{center}
\end{figure}

The calculated neutron $^3PF_2$ gaps for the $\beta$-stable matter are depicted in
Figure~\ref{fig:n3PF2-beta}, with $Z=1$ (upper ponel) and  $Z \neq 1$  (lower panel).
Similar to the case of the pure neutron matter (Dong et al. 2013), the $Z$ factor
effect quenches the peak value by about one order of magnitude, and it is extremely
sizable at higher densities. Ding et al. (2015) calculated the influence of
short-range correlations on the $^3PF_2$ pairing gap in pure neutron matter
at high density with a different method, and also found that the gap is strongly
suppressed. The peak value drops to 0.04 MeV and the superfluidity domain
shrinks to $0.1-0.4$ fm$^{-3}$. Such a weak superfluidity is not expected to explain the
observed rapid cooling of NSs in the case of Cas A via the enhanced neutrino emission
from the onset of the breaking and formation of neutron $^3PF_2$ Cooper pairs. Therefore, the role
of the $^3PF_2$ superfluidity in NS cooling is limited. The $^3PF_2$ pairing gaps do not change very much,
even if the fraction of each component is controlled by a soft symmetry energy, such as in the case
of APR EOS (Akmal et al. 1998).

\begin{figure}[htbp]
\begin{center}
\includegraphics[width=0.45\textwidth]{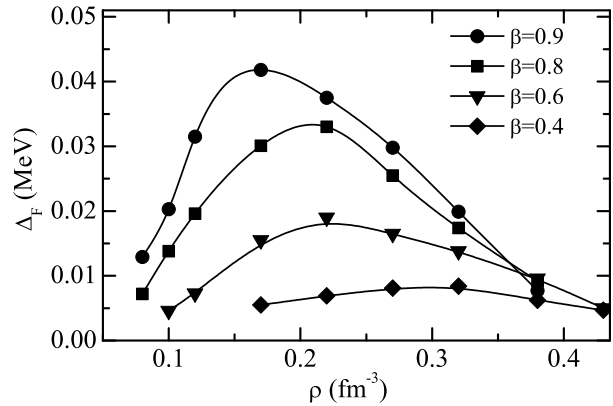}
\caption{Neutron $^3PF_2$ gap as a function of nucleonic density in asymmetric
matter for various isospin asymmetries $\beta$.}\label{fig:n3PF2}
\end{center}
\end{figure}

To further explore the $^3PF_2$ superfluidity, the neutron gaps as a function of the total
nucleonic density $\rho$ for different isospin asymmetries $\beta$ were calculated, and the results are
shown in Figure~\ref{fig:n3PF2}. The isospin asymmetry is defined as $\beta=(\rho_n-\rho_p)/(\rho_n+\rho_p)$, where
$\rho_n$ and $\rho_p$ are the neutron and proton density, respectively. As $\beta$ increases, the neutron density for a given total
nucleonic density increases either. Accordingly, the peak value of the gap increases and their
corresponding location shifts to lower density, that is primarily attributed to the
decrease of the neutron $Z$ factor for increasing $\beta$. Anyway, due to the departure of the
system from the ideal degenerate limit, the neutron $^3PF_2$ gap is dramatically suppressed
and the peak value is no more than 0.05 MeV. For the nuclear matter with small isospin asymmetry,
the gap may be vanishingly small. Therefore, in finite nuclei, the pairing comes from
the $^3PF_2$ channel is completely negligible.

\subsection{Neutrino emission}\label{intro}\noindent

Since the deformation of the Fermi surface hinders particle-hole excitations around
the Fermi level, the neutrino emission is accordingly expected to be
depressed. To determine the neutrino emissivity, the key step is to derive the
nucleon momentum distribution $n(k)$
\begin{equation}
n(k)=\int \frac{d\omega }{2\pi }S(k,\omega )f(\omega ),
\end{equation}
at finite temperature $T$ and $Z$ (Kadanoff \& Baym 1962).

The spectral function $S(k,\omega )$ is a weighting function with
total weight unity. $f(\omega )=1/\left[ 1+\exp (\frac{\omega -\mu
}{T})\right] $ is Fermi distribution with temperature $T$ and
chemical potential $\mu$. In the limit when $k$ is extremely close
to the Fermi momentum, the spectral function $S(k,\omega )$ can be
expressed as
\begin{equation}
S(k,\omega )\approx Z(k)\delta (\omega -\omega _{F}),k\approx k_{F}
\end{equation}
As a consequence, the momentum distribution function $n(k)$ of
nucleon close to the Fermi surface is given as
\begin{equation}
n(k)=\frac{Z_F}{1+\exp (\frac{\omega -\mu }{T})},k\approx k_{F}.\label{A}
\end{equation}

The most efficient neutrino emission is provided by DU
processes in the NS core
\begin{equation}
n\rightarrow p+l+\overline{\nu _{l}}, p+l\rightarrow n+\nu _{l},
\end{equation}
which correspond to neutron beta decay and proton electron capture, respectively.
The DU process occurs only if the proton fraction is sufficiently high.
We derive the neutrino emissivity $Q^{(D)}$ of the DU process under
the beta-equilibrium condition with the inclusion of the $Z$ factor. It  is given by
\begin{equation}
Q^{(D)}=2\int \frac{d\bm{k}_{n}}{(2\pi )^{3}}\varepsilon _{\nu
}dW_{i\rightarrow f}n_{n}(k)\left[ n_{p}(k)|_{T=0}-n_{p}(k)\right]
\left( 1-f_{l}\right) ,\label{B}
\end{equation}
where $dW_{i\rightarrow f}$ is the beta decay differential probability, $n_i(k)$ is
the distribution function of nucleons including the $Z$ factor, and $f_l$
the Fermi-Dirac distribution function of leptons. Since the main contribution to this integral
stems from the very narrow regions of momentum space close to the Fermi surface
of each particle, the distribution function of Equation (\ref{A}) can be employed here, and thus
the above Equation (\ref{B}) reduces to
\begin{eqnarray}
Q^{(D)} &\approx &2\int \frac{\varepsilon _{\nu }d\bm{k}_{n}}{(2\pi )^{3}%
}dW_{i\rightarrow f}\frac{Z_{F,n}}{1+\exp (\frac{\omega _{n}-\mu _{n}}{T})}%
\frac{Z_{F,p}}{1+\exp (\frac{-\omega +\mu _{p}}{T})}\left( 1-f_{l}\right)
\nonumber \\
&=&2Z_{F,n}Z_{F,p}\int \frac{d\bm{k}_{n}}{(2\pi )^{3}}\varepsilon
_{\nu }dW_{i\rightarrow f}f_{n}\left( 1-f_{p}\right) \left(
1-f_{l}\right)
\nonumber \\
&=&Z_{F,n}Z_{F,p}Q_{0}^{(D)},
\end{eqnarray}
where $Q_{0}^{(D)}$ is the neutrino emissivity of the DU process without
introducing the $Z$ factor, which has been well studied (see,e.g.,
the review papers, Yakovlev et al. 1999, 2001).

\begin{figure*}[htbp]
\begin{center}
\includegraphics[width=0.95\textwidth]{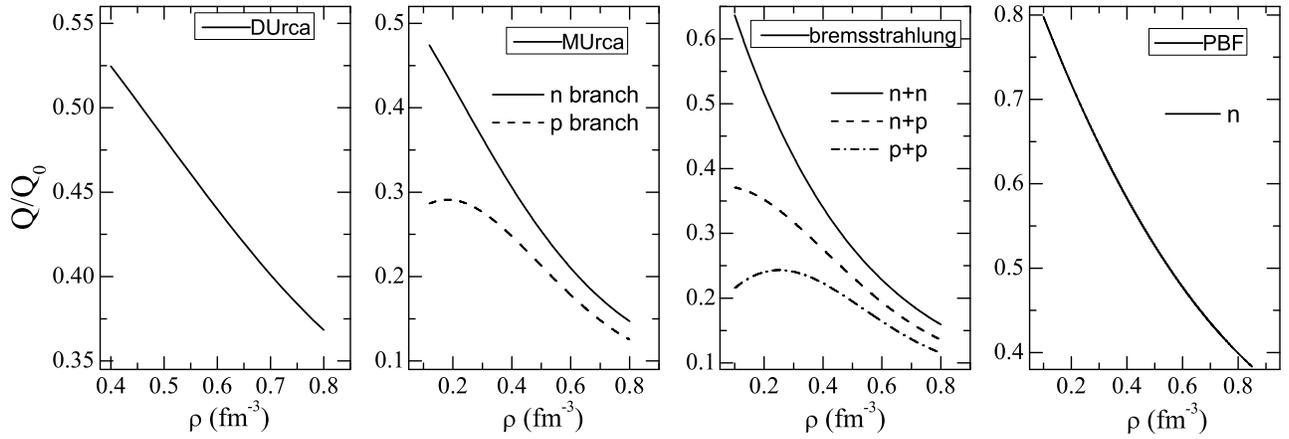}
\caption{$Q/Q_0$ for the DU, MU, nucleon-nucleon bremsstrahlung and
PBF processes vs density $\rho$ in $\beta$-stable NS
matter. The fraction of each component is determined by symmetry
energy from the Brueckner theory.}\label{fig:emission}
\end{center}
\end{figure*}

It is usually believed that one of the main neutrino energy loss
processes in NSs are the MU process, which
are several orders of magnitude less efficient than the DU processes.
The MU processes differ from the direct one by a bystander nucleon
required to allow momentum conservation
\begin{equation}
 n+N\rightarrow p+N+l+\overline{\nu _{l}},p+N+l\rightarrow n+N+\nu _{l},
\end{equation}
where $N$ denotes neutron or proton. It is labeled by the
superscript $MN$, where $N=n(p)$ denotes the neutron (proton) branch
of the MU processes. Analogously to the procedure for the DU
processes, we derive the neutrino emissivity under the condition of
beta equilibrium
\begin{eqnarray}
Q^{(Mn)} &=&Z_{F,n}^{3}Z_{F,p}Q_{0}^{(Mn)},  \nonumber \\
Q^{(Mp)} &=&Z_{F,n}Z_{F,p}^{3}Q_{0}^{(Mp)}.
\end{eqnarray}

The nucleon-nucleon bremsstrahlung process
\begin{equation}
N+N\rightarrow N+N+\nu +\overline{\nu },
\end{equation}
is another neutrino process based on neutral current that produces
$\nu \overline{\nu }$ pairs to take away the neutron star thermal
energy, but is much less efficient than the MU processes. The
corresponding neutrino emissivities with the inclusion of the $Z$
factor are
\begin{eqnarray}
Q^{(nn)} &=&Z_{F,n}^{4}Q_{0}^{(nn)},  \nonumber \\
Q^{(np)} &=&Z_{F,n}^{2}Z_{F,p}^{2}Q_{0}^{(np)},  \nonumber \\
Q^{(pp)} &=&Z_{F,p}^{4}Q_{0}^{(pp)}.
\end{eqnarray}

The onset of pairing also opens a new channel of neutrino emission due to
the continuous formation and breaking of Cooper pairs
\begin{equation}
N+N\rightarrow \left[ NN\right] +\nu +\overline{\nu },
\end{equation}
that leads to energy release via $\nu \overline{\nu }$ emission,
being analogous to the nucleon-nucleon bremsstrahlung process. It is
very intense at temperatures slightly below the critical temperature
$T_c$ (Page et al. 2006, 2009), and it can be one order of magnitude more
efficient than the pairing unsuppressed MU processes (Page et al. 2006).
This Cooper pair breaking and formation (PBF) process due to
neutron $^3PF_2$ pairing may be employed to explain the observed
rapid cooling of the neutron star in Cas A (Page et al. 2011). With
the inclusion of $Z$ factor, the neutrino emissivity $Q$ is given as
\begin{eqnarray}
Q^{(PBF,n)} &=&Z_{F,n}^{2}Q_{0}^{(PBF,n)}.
\end{eqnarray}

Figure~\ref{fig:emission} displays the calculated $Q/Q_0$ for the
DU, MU, and bremsstrahlung processes as a function of density
$\rho$. The DU process occurs just when the proton fraction exceeds
a critical threshold, and the threshold relies on the density
dependence of the symmetry energy at high densities. A stiff
symmetry energy, such as that from the Brueckner theory, gives a low
threshold. Frankfurt {\it et al.} (2008) showed that the
modification of the nucleon momentum distribution due to the
short-range correlations results in a significant enhancement of the
neutrino emissivity of the DU process, and the DU process has a
probability of being opened even for low proton fractions. However,
our calculations reveal that the neutrino emissivity is reduced by
more than $\simeq 50\%$, that is in complete contrast to their
conclusion. We stress that, the short range correlations responsible
of the Fermi surface depletion should be embodied in the
quasiparticle properties, because the real system is actually a
degenerate system of quasiparticles with the same Fermi momentum as
the previous one. Therefore the kinematic conditions giving rise to
threshold for DU process does not change. The beta decay of neutrons
and its inverse reaction can be cyclically triggered just by thermal
excitation, and thus $k\gg k_F$ and $k\ll k_F$ states do not
participate into the DU process because the thermal energy $\sim k_B
T$ is too low to excite those states. Accordingly, the occupation
probability of proton hole for the thermal excitation, namely
$[n_{p}(k)|_{T=0}-n_{p}(k)]$, should be employed in Equation
(\ref{B}), instead of $[1-n_{p}(k)]$. The neutrino emissivity for
the MU processes are reduced by $> 50\%$ for the neutron branch and
$> 70\%$ for the proton branch. Because the proton $Z$ factor is
lower than the neutron one, as shown in Figure~\ref{fig:Z}, the
$Q^{MN}$ of proton branch is more intensively depressed by the Fermi
surface depletion. At high densities such as $\rho=0.8$ fm$^{-3}$,
the $Q^{MN}$ is reduced by one order of magnitude. Similarly, the
nucleon-nucleon bremsstrahlung for $n-n$, $n-p$ and $p-p$ are
reduced more distinctly at high densities. The computed $Q/Q_0$ for
those processes do not rely on whether or not the superfluidity set
on. The $Q/Q_0$ with a soft symmetry energy from APR (Akmal et al.
1998) does not provide very different results compared with those
within BHF for the MU, nucleon-nucleon bremsstrahlung and PBF
processes, but the threshold of the DU process is much larger due to
the softer symmetry energy.

\subsection{Heat capacity}\label{intro}\noindent

The specific heat capacity of the stellar interior is the
sum of the contributions of each fraction $i$ (leptons and nucleons) (Page et al. 2004)
\begin{equation}
c_{v}=\underset{i}{\sum }c_{v,i},\text{\ \ with }c_{v,i}=\left( \frac{%
m_{i}^{\ast }p_{F,i}}{3\hbar ^{3}}\right) k_{B}^{2}T.
\end{equation}
$m_i^{\ast }$ and $p_{,F,i}$ are effective mass and Fermi momentum of particle $i$.
This equation is obtained under the assumption of non-correlated gas described by the Fermi-Dirac distribution.
If the effect of the Fermi surface depletion is included, the specific heat of the nucleon$i$ is given by
\begin{equation}
c_{v,i}=Z_{F,i}\left( \frac{m_{i}^{\ast }p_{F,i}}{3\hbar ^{3}}\right)
k_{B}^{2}T,\text{ \ }i=n,p.
\end{equation}
The heat capacity can be still altered by strong superfluidity.

\begin{figure}[htbp]
\begin{center}
\includegraphics[width=0.45\textwidth]{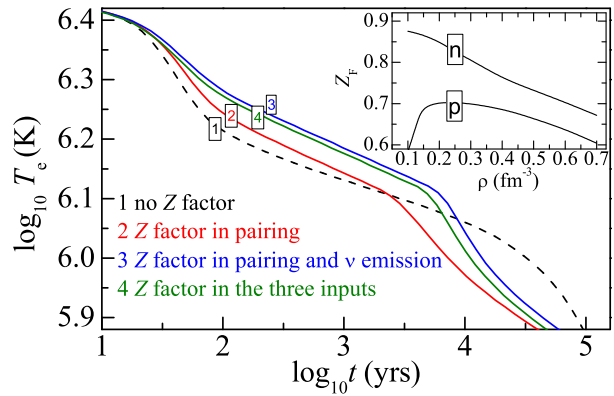}
\caption{Cooling curves of a canonical NS. The stellar structure is built with the APR EOS.
The calculations without any $Z$ factors, with $Z$ factors only in superfluidity, with $Z$ factors both in superfluidity
and neutrino emission and with $Z$ factors in the three inputs are shown for comparison. The inset shows the
neutron and proton $Z$ factors vs density in $\beta$-stable matter,  the
fraction of each component being  determined by the APR EOS. (A color version of this figure is available in the online journal.)
}\label{fig:cooling}
\end{center}
\end{figure}

\section{Neutron star cooling}\label{intro}\noindent

To show the influence of the $Z$-factor induced
reduction of the above three inputs afore discussed, namely, pairing gaps, neutrino emissivity and heat capacity,
on NS cooling, we calculated the cooling curve for canonical neutron stars
using the publicly available code NSCool$^{[1]}$ \footnotetext[1]{http://www.astroscu.unam.mx/neutrones/home.html}. We
employ the minimal cooling paradigm (Page et al. 2006) without fast
neutrino emission, with no charged meson condensate, no hyperons, no
confinement quarks in canonical NSs. Therefore, we select
the APR EOS and the above results as inputs, and the cooling curves are displayed in Figure~\ref{fig:cooling}.

The $Z$ factor suppressing the neutron $^3PF_2$ superfluidity, retards the neutron star cooling for
the first $3\times 10^3$ years but accelerates the cooling thereafter.
The critical temperature $T_c$ for neutron $^3PF_2$ superfluidity is quite low as a result of the inclusion of $Z$ factor.
Therefore, the PBF cooling channel opens as soon as the stellar tempereture falls below the $T_c$,
which leads to a relatively fast cooling because it is more efficient than MU processes.
The weak neutron $^3PF_2$ superfluidity, dramatically quenched by the Fermi surface
depletion, cannot play a significant role in cooling of young NSs. The $Z$ factor substantially reduces
the neutrino emission of MU, nucleon-nucleon bremsstrahlung and PBF processes,
and thus it slows down the thermal energy loss. Therefore, it significantly retards the cooling.
The heat capacity of the neutron star is reduced by the $Z$ factor, with the result that the
thermal energy turns lower than in the case of  excluding of the $Z$-factor effect. As a consequence,
the NS cooling turns out to be enhanced, but this effect is not so sharpt, as shown in Figure~\ref{fig:cooling}.
It is well-known that the neutrino emission and heat capacity are sensitive to the superfluidity of stellar interior.
However, due to the weak neutron $^3PF_2$ superfluidity  reduced by the $Z$ factor,
neutrino emissivity and heat capacity cannot be suppressed. In a word, the effect of the Fermi surface depletion
of nucleons on NS  cooling cannot be neglected, when an  accurate theoretical study of the cooling is performed, as in the case of the
cooling of the NS remnant of Cas A.

\section{Summary}\label{intro}\noindent

We have investigated the superfluidity, neutrino emissivity for DU,
MU, nucleon-nucleon bremsstrahlung and PBF processes and heat capacity,
taking into account the Fermi surface depletion characterized by the
$Z$ factor. The $Z$ factor is calculated in the framework of the BHF approach
using the two body AV18 force plus a microscopic three body force. The superfluidity, neutrino emissivity
and heat capacity are reduced by the $Z$ factor, specially at high baryonic density.
The effect of the Fermi surface depletion is needed to be included in a  theoretically rigorous
exploration of the NS thermal evolution, such as for the NS remnant in Cas A whose
cooling rate was measured. Finally, based on the above results, we calculated the
cooling curve for canonical NSs  using the APR EOS, and we found that the Fermi surface depletion visibly affects the NS cooling.

\section*{Acknowledgement}

\label{intro}\noindent
This work was supported by the National Natural Science
Foundation of China under Grants No. 11405223, No.
11175219, No. 11275271, No. 11435014 and No. 11175074, by the 973
Program of China under Grant No. 2013CB834405, by the
Knowledge Innovation Project (KJCX2-EW-N01) of Chinese
Academy of Sciences, by the Funds for Creative Research
Groups of China under Grant No. 11321064, and by the Youth
Innovation Promotion Association of Chinese Academy of
Sciences.



\begin{thebibliography}{MBT}

Akmal, A., Pandharipande, V. R., \& Ravenhall, D. G. 1998, PhRvC, 58, 1804 \\

Baldo, M., Cugnon, J., Lejeune, A., \& Lombardo, U. 1992, NuPhA, 536, 349 \\

Blaschke, D., Grigorian, H., Voskresensky, D. N., \& Weber, F. 2012, PhRvC, 85, 022802(R)\\

Blaschke, D., Grigorian, H., \& Voskresensky, D. N. 2013, PhRvC, 88, 065805\\

Bonanno, A., Baldo, M., Burgio, G. F., \& Urpin, V. 2014, A\&A, 561, L5 \\

Day, B. D. 1978, Rev. Mod. Phys., 50, 495\\

Ding, D., Rios, A., Dickhoff, W. H., Dussan, H., Polls, A., \& Witte, S. J. 2015, arXiv:1502.05673v1\\

Dong, J. M., Lombardo, U., \& Zuo, W. 2013, PhRvC, 87, 062801(R)\\

Frankfurt, L., Sargsian, M., \& Strikman, M. 2008, IJMPA, 23, 2991 \\

Haensel, P., Potekhin, A. Y., \& Yakovlev, D. G. 2006, {\it Neutron Stars 1},
(Springer)\\

Heinke, C. O., \& Ho, W. C. G. 2010, ApJL, 719, L167 \\

Jeukenne, J. P., Lejeune, A., \& Mahaux, C. 1976, PhR, 25, 83 \\

Kadanoff, L. P., \& Baym, G. 1962, Quantum Statistical Mechanics, (New
York) \\

Li, Z. H., Lombardo, U., Schulze, H.-J., \& Zuo, W. 2008, PhRvC, 77, 034316 \\

Migdal, A. B. 1957, Sov. Phys. JETP, 5, 333; Luttinger, J. M. 1960,
PhRv, 119, 1153 \\

Migdal, A. B. 1967, Theory of Finite Fermi Systems and Applications
to Atomic Nuclei (Interscience, London) \\

Newton, W. G., Murphy, K., Hooker, J., \& Li, B.-A. 2013, ApJL, 779, L4 \\

Page, D., Lattimer, J. M., Prakash, M., \& Steiner, A. W. 2004, ApJS, 155, 623  \\

Page, D., Geppert, U., \& Weber, F. 2006, NuPhA, 777, 497 \\

Page, D., Lattimer, J. M., Prakash, M., \&
Steiner, A. W. 2009, ApJ, 707, 1131  \\

Page, D., Prakash, M., Lattimer, J. M., \& Steiner, A. W. 2011, PhRvL, 106, 081101 \\

Posselt, B., Pavlov, G. G., Suleimanov, V., \& Kargaltsev, O. 2013, ApJ, 779, 186 \\

Sedrakian, A 2013, A\&A, 555, L10 \\

Shternin, P. S., Yakovlev, D. G., Heinke, C. O., et al. 2011, MNRAS, 412, L108 \\

Yakovlev, D. G., Levenfish, K. P., Shibanov, Y. A. 1999, Physics Uspekhi, 42, 737 \\

Yakovlev, D. G., Kaminker, A. D., Gnedin, O. Y., \& Haensel, P. 2001, PhR, 354, 1 \\

Yakovlev, D. G., \& Pethick, C. J. 2004, Annu. Rev. Astron. Astrophys. 42, 169 \\


\end{thebibliography}
\end{document}